\documentclass{PoS}

\def\be{\begin{equation}}
\def\ee{\end{equation}}
\def\bea{\begin{eqnarray}}
\def\eea{\end{eqnarray}}

\title{Production of forward di-jet in p+Pb collisions in the
    small-$x$ improved TMD factorization framework}

\ShortTitle{Forward di-jet}

\author{Krzysztof Kutak\\       
Institute of Nuclear Physics, Polish Academy of Sciences,\\
{\small\it Radzikowskiego 152, 31-342 Krak\'ow, Poland}\\
        E-mail: \email{Krzysztof.Kutak@ifj.edu.pl}}

\abstract{We report on recent study of the production of forward di-jets in proton-proton and proton-lead collisions at the Large Hadron Collider
with Improved Transversal Momentum Factorization \cite{vanHameren:2016ftb}. The results as compared to results obtained within High Energy Factorization show noticable effects
related to detailed treatment of nonlinear effects.}

\FullConference{38th International Conference on High Energy Physics\\
		3-10 August 2016\\
		Chicago, USA}

\begin{document}
\section{Introduction}
Measurements of forward jet or particle production in high-energy hadronic collisions provide unique
opportunities to study the QCD dynamics \cite{Deak:2009xt,Deak:2009ae,Deak:2010gk,Deak:2011ga,Deak:2011gj,Kutak:2012rf} and in particular of the non-linear parton saturation regime~\cite{Marquet:2007vb,Gribov:1984tu}.
Such processes, in which, for kinematical reasons, high-momentum partons from one of the colliding hadrons mainly scatter with small-momentum partons from the other, are called dilute-dense collisions. Indeed, the density of the large-$x$ partons in the projectile hadron is small, while the density 
of the small-$x$ gluons in the target hadron is large, and the former, well understood in perturbative QCD, can be used to probe the dynamics of the latter. This is true already in proton-proton collisions, although using a target nucleus does enhance the dilute-dense asymmetry of such collisions.
 
RHIC measurements have provided some evidence for the presence of saturation effects
in the data, the most compelling of which is the successful description of forward di-hadron
production \cite{Albacete:2010pg,Stasto:2011ru,Lappi:2012nh}, using the most up-to-date
theoretical tools available at the time in the Color Glass Condensate (CGC) framework \cite{Gelis:2010nm,Albacete:2014fwa}. In particular, this 
approach predicted the suppression of azimuthal correlations in d+Au collisions compared to p+p collisions \cite{Marquet:2007vb}, which was observed later experimentally \cite{Adare:2011sc,Braidot:2010zh}.

In this context, we shall consider forward di-jet production in proton-lead versus proton-proton collisions. In that case, it was shown in \cite{Kotko:2015ura} that the full complexity of the CGC machinery is not needed. Indeed, for the di-hadron process at RHIC energies, no particular ordering of the momentum scales involved is assumed in CGC calculations, while at the LHC one can take advantage of the presence of final-state partons with transverse momenta much larger than the saturation scale to obtain simplifications. On the flip side, different complications - left for future studies - are expected to arise due to QCD dynamics relevant at large transverse momenta and not part of the CGC framework, such as Sudakov logarithms \cite{Mueller:2012uf,Mueller:2013wwa,vanHameren:2014ala,Kutak:2014wga} or coherence in the QCD evolution of the gluon density \cite{Ciafaloni:1987ur,Catani:1989sg,Catani:1989yc}.

The goal of this article is to report on application \cite{vanHameren:2016ftb} of that new formulation, dubbed improved TMD (ITMD) factorization which is a generalization for considered process of High Energy Factorization \cite{Deak:2009xt,Catani:1990eg} and TMD factorization \cite{Angeles-Martinez:2015sea}.
By comparing the forward di-jet production cross sections in proton-lead and proton-proton collisions, we can clearly see the onset of parton 
saturation effects, as we go from a kinematical regime in which $k_t\sim P_t$ towards one where $k_t\sim Q_s$, and we obtain a good estimation of the size of those effects where they are the biggest, which is for nearly back-to-back jets. We note that probing non-linear effects of similar strength with single-inclusive observables requires to make the only transverse momentum involved in those processes of the order of the saturation scale, which may not be easy experimentally. With di-jets, assuming $P_t\sim 20$ GeV and $k_t\sim Q_s \sim 2$ GeV, we can reach $R_{pPb}\sim 0.5$.

\section{The ITMD factorization formula for forward di-jets in dilute-dense collisions}

We consider the process of inclusive forward di-jet production in hadronic collisions
\begin{equation}
  p (p_p) + A (p_A) \to j_1 (p_1) + j_2 (p_2)+ X\ ,
\end{equation}
where the four-momenta of the projectile and the target are massless and purely longitudinal.
The longitudinal momentum fractions of the incoming parton from the projectile, $x_1$, and the gluon from the target, $x_2$, can be expressed in
terms of the rapidities $(y_1,y_2)$ and transverse momenta $(p_{t1},p_{t2})$ of the produced jets as
\be
x_1  = \frac{p_1^+ + p_2^+}{p_p^+}   = \frac{1}{\sqrt{s}} \left(|p_{1t}|
e^{y_1}+|p_{2t}| e^{y_2}\right)\ , \quad
x_2  = \frac{p_1^- + p_2^-}{p_A^-}   = \frac{1}{\sqrt{s}} \left(|p_{1t}| e^{-y_1}+|p_{2t}| e^{-y_2}\right)\ .
\ee
By looking at jets produced in the forward direction, we effectively select those fractions to be $x_1 \sim 1$ and $x_2 \ll 1$.
Since the target A is probed at low $x_2$, the dominant contributions come from the subprocesses in which the incoming parton on the target side is a gluon
\begin{equation}
  qg  \to  qg\ ,
  \qquad \qquad 
  gg  \to  q\bar q\ ,
  \qquad \qquad 
  gg  \to  gg\ .
\label{eq:3channels}
\end{equation}

Moreover, the large-$x$ partons of the dilute projectile are described in terms of the usual parton distribution functions of collinear factorization $f_{a/p}(x_1)$ while the small-$x$ gluons of the dense target are described by TMD distributions $\Phi_{g/A}(x_2,k_t)$. Indeed, the momentum of the incoming gluon from the target is not only longitudinal but also has a non-zero transverse component of magnitude
\be
k_t = |p_{1t}+p_{2t}|
\label{eq:ktglue}
\ee
which leads to imbalance of transverse momentum of the produced jets: $k_t^2 =|p_{1t}|^2 + |p_{2t}|^2 + 2|p_{1t}||p_{2t}| \cos\Delta\phi$.
The validity domain of ITMD factorization is
\be
Q_s(x_2)\ll P_t
\ee
where $P_t$ is the hard scale of the process, related to the individual jet momenta $P_t \sim |p_{1t}|,|p_{2t}|$. By contrast, the value of $k_t$ can be arbitrary.

The ITMD factorization formula reads \cite{Kotko:2015ura}
\begin{equation}
\frac{d\sigma^{pA\rightarrow {\rm dijets}+X}}{d^{2}P_{t}d^{2}k_{t}dy_{1}dy_{2}}=\frac{\alpha_{s}^{2}}{(x_1 x_2 s)^{2}}
\sum_{a,c,d} \frac{x_1 f_{a/p}(x_1)}{1+\delta_{cd}}\sum_{i=1}^{2}K_{ag^*\to cd}^{(i)}(P_t,k_t)\Phi_{ag\rightarrow cd}^{(i)}(x_2,k_t)\ .
\label{eq:itmd}
\end{equation}
It involves several gluon TMDs $\Phi_{ag\rightarrow cd}^{(i)}$ (2 per channel), with different operator definitions, that are accompanied by different hard factors $K_{ag^*\to cd}^{(i)}$. Those where computed in \cite{Kotko:2015ura} using either Feynman diagram techniques, or color-ordered amplitude methods. They encompass the improvement over the TMD factorization formula derived in Ref.~\cite{Dominguez:2011wm} where the matrix elements were on-shell and a function of $P_t$ only.

We would like to point out that the ITMD factorization formula \ref{eq:itmd} was build in order to contain both the HEF and the TMD expressions as its limiting cases, and as such should be considered no more than an interpolating formula. We note however, that if one would be able to directly derive a factorization formula valid for $Q_s\ll P_t$ regardless of the value of $k_t$, any additional term compared to \ref{eq:itmd} should vanish in both limits $Q_s\sim k_t\ll P_t$ and $Q_s\ll k_t\sim P_t$.

\section{Numerical studies of the forward di-jet cross section}
\label{sec:Num_studies}
We move now to the numerical results\footnote{the calculations were performed using Monte Carlo programs \cite{vanHameren:2016kkz,KotkoLxJet}} for forward di-jet production in p+p and p+Pb collisions at the LHC.
We consider a center-of-mass energy of $8.16\, \mathrm{TeV}$, and generate all our predictions with the
forward region defined as the rapidity range $3.5 < y < 4.5$ on one side of the detector. The two hardest
jets are required to lie within this region and we also impose a cut on the minimal transverse momentum
of each two jets: $p_{t0}=20\, \mathrm{GeV}$. In such a setup, the cross section still may be divergent due
to collinear singularities. These are cut-off by applying a jet algorithm on the final state momenta with a
delta-phi-rapidity cut $R=0.5$. Finally, we require the jets to be ordered according to increasing transverse
momentum, that is we have $|p_{t1}|>|p_{t2}|>p_{t0}$.
For the collinear parton distributions that enter the ITMD formula, we chose the general-purpose CT10 set.
For the central value of the factorization and renormalization scale, we choose the average transverse momentum of the
two leading jets, $\mu_F=\mu_R=\frac12 (|p_{t1}|+|p_{t2}|)$. We will produce error bands corresponding to the renormalization
and factorization scale uncertainties by varying the central numbers from half to twice their value.

For the various observables  ${\cal O}$ shown below, we also consider the nuclear modification factors defined as
\begin{equation}
R_{\rm pPb} = \frac{\displaystyle \frac{d\sigma^{p+Pb}}{d{\cal O}}}
                 {\displaystyle A\ \frac{d\sigma^{p+p}}{d{\cal O}}}\,.
\label{eq:RpA}
\end{equation}
with $A=208$ for Pb. In our approach, in the absence of saturation effects, or in the case in which they are
equally strong in the nucleus and in the proton, this ratio is equal to unity. If,
however, the non-linear evolution plays a more important role in the case of the
nucleus, the $R_{\rm pPb}$ ratio will be suppressed below 1.

\begin{figure}[t]
  \begin{center}
\includegraphics[width=.48\textwidth]{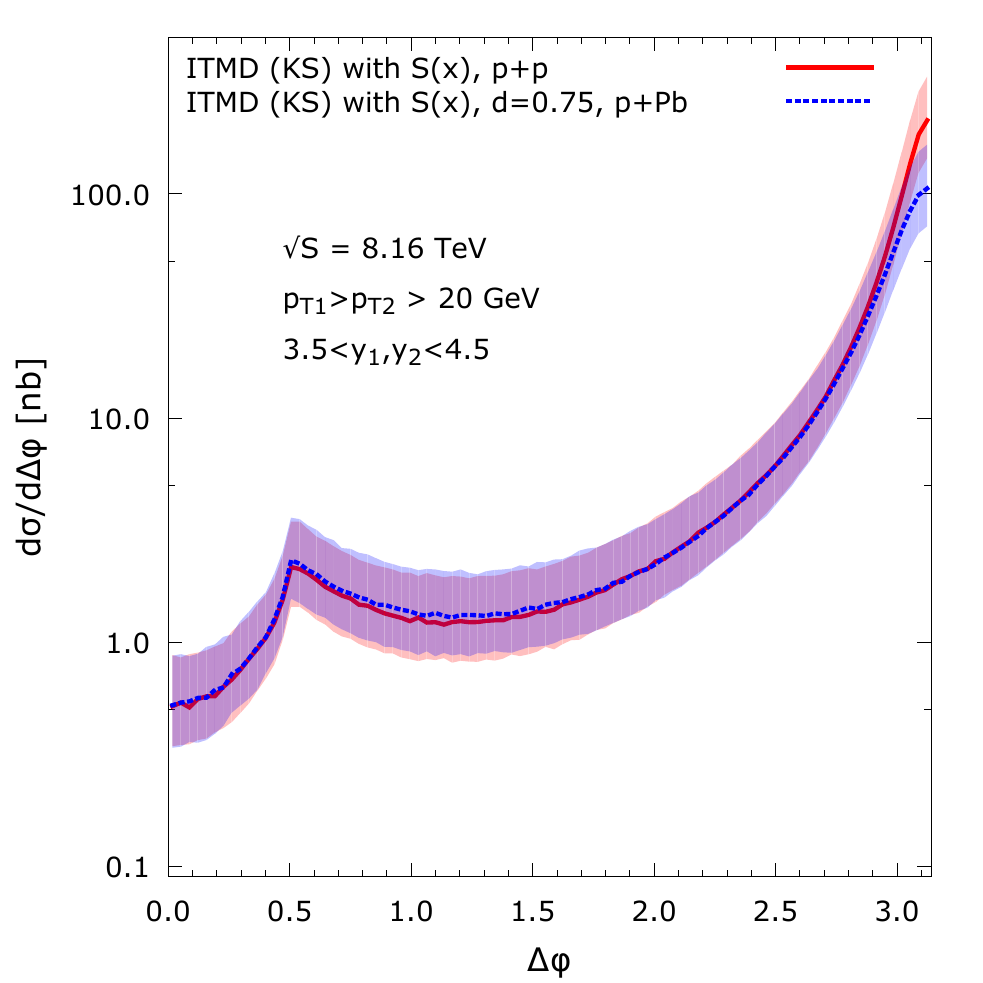}
    \includegraphics[width=0.48\textwidth]{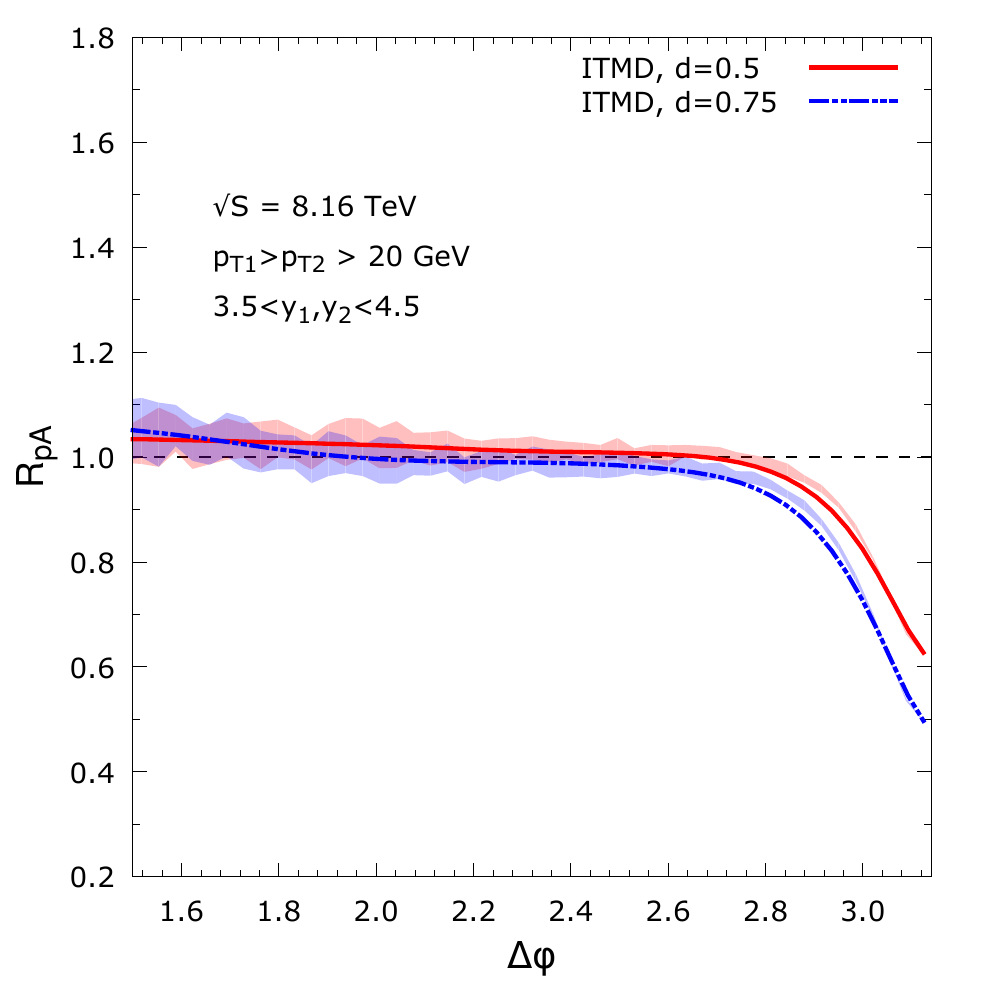}
  \end{center}
  \caption{Left plot: differential cross section as a function of the azimuthal angle between the jets for
  p+p and p+Pb collisions (rescaled by the number of nucleons). The distributions are identical
  everywhere expect near $\Delta\phi\simeq\pi$, where saturation are the strongest. Right plot:
  nuclear modification factors for two values of the nuclear saturation scale, providing an uncertainty band.}
  \label{fig:dphipvsPb}
\end{figure}

We start by investigating the azimuthal correlations, with the azimuthal angle between the jets $\Delta\phi$ defined to lie within $0<\Delta\phi<\pi$.

In Fig.~\ref{fig:dphipvsPb}  we compare the $\Delta\Phi$ distribution in p+p and p+Pb collisions. After rescaling the p+Pb cross section by the number of nucleons, we obtain identical distributions almost everywhere. It is only for nearly back-to-back jets, around $\Delta\phi\simeq\pi$, that saturation effects induce a difference. This difference is better appreciated on the nuclear modification factor, which goes from unity to 0.6, as $\Delta\phi$ varies from $\sim2.7$ to $\pi$. Two values of the parameter $c$ have been considered, which makes up an uncertainty band that turns out to be rather small. This means that the uncertainty related to the value of the saturation scale of the lead nucleus does not strongly influence the predicted $R_{\rm pPb}$ suppression.

\begin{figure}[t]
  \begin{center}
    \includegraphics[width=0.48\textwidth]{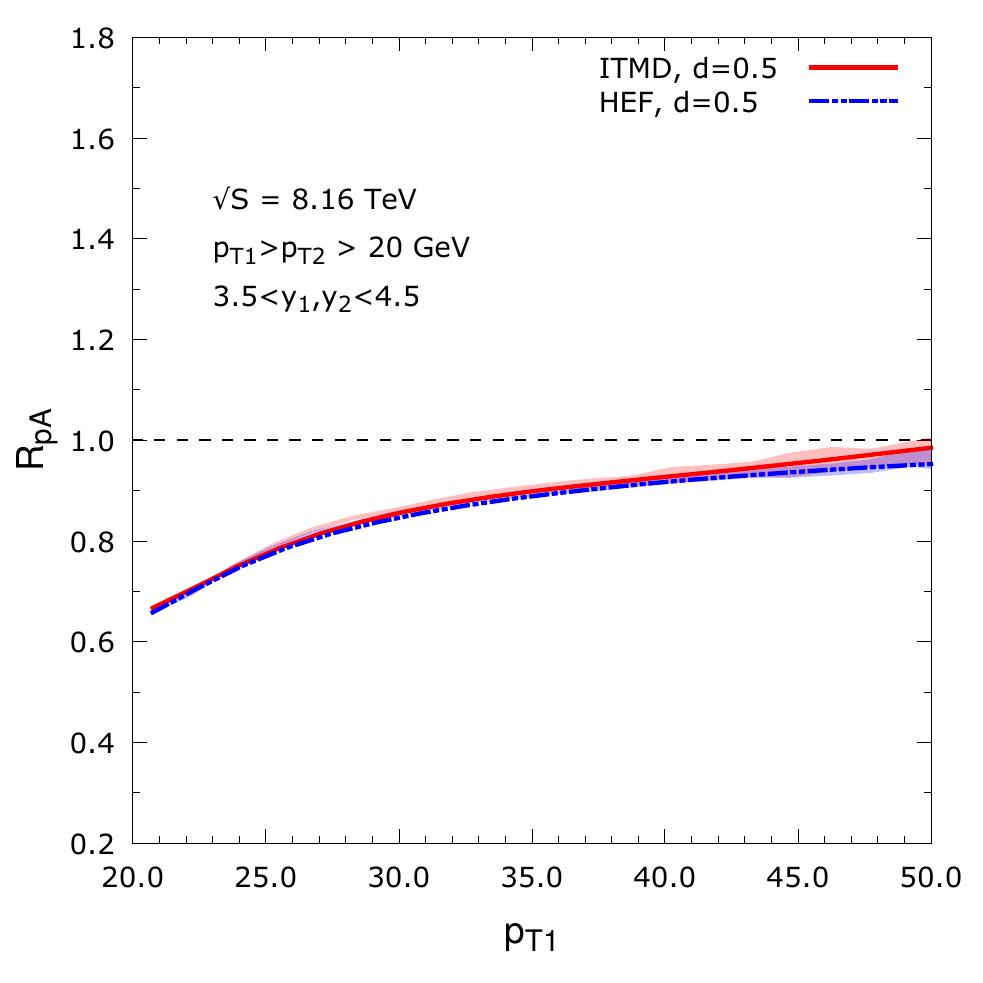}
    \includegraphics[width=0.48\textwidth]{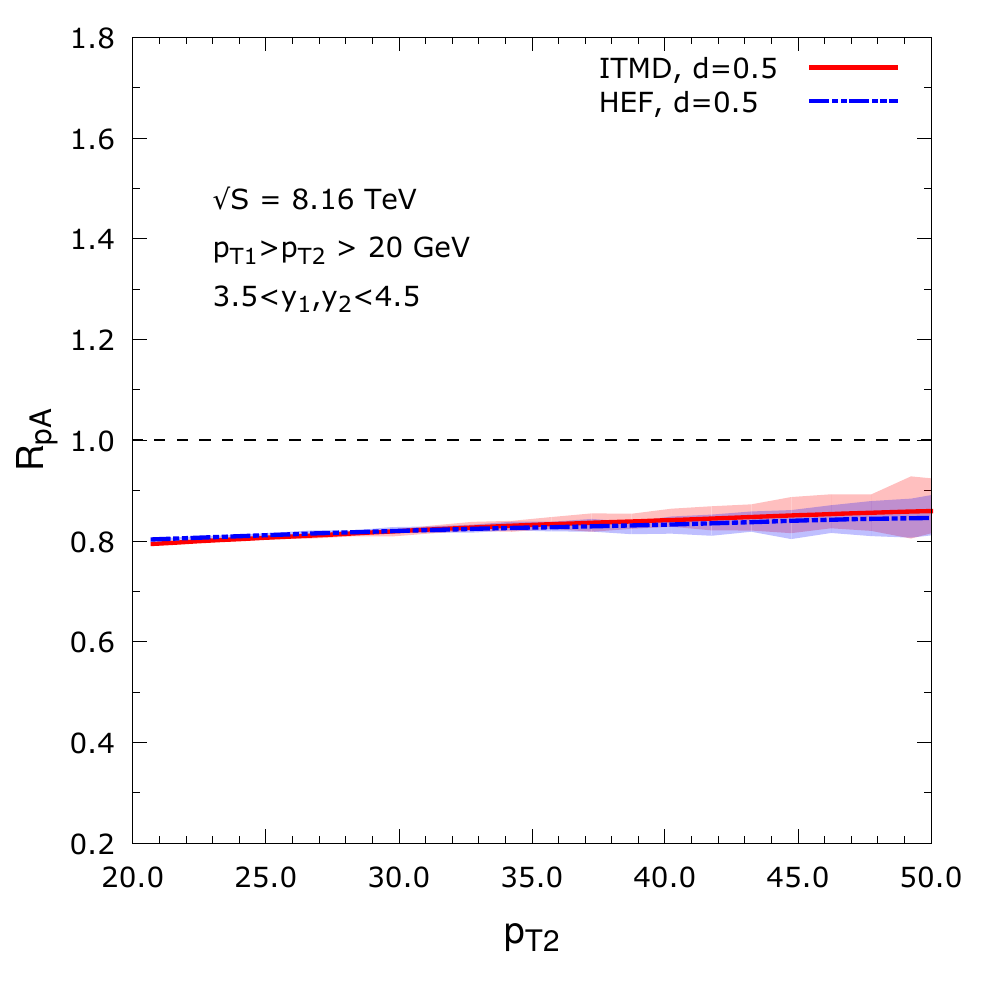}
  \end{center}
  \caption{Nuclear modification factors as a function of the transverse momentum of the leading (left) and subleading (right) jet, 
   comparing the new ITMD approach with previously obtained HEF results.}
  \label{fig:ptHEF-ITMD}
\end{figure}

Finally, in Fig.~\ref{fig:ptHEF-ITMD} we display the nuclear modification factors as a function
of the transverse momentum of the leading and sub-leading jet. Our conclusions are similar for these observables: the new ITMD
predictions are similar to the previously obtained HEF results, due to the fact that the ITMD/HEF ratio is similar in p+p and p+Pb collisions.
This means that the HEF framework, which is incorrect for nearly back-to-back jets - since in this formalism all the gluon TMDs are considered
equal regardless of the kinematics - can nevertheless be safely used for $R_{\rm pPb}$ calculations.

\section{Conclusions}
  
In this paper, we have studied forward di-jet production in proton-proton and proton-lead collisions,
using the small-x improved TMD factorization framework Eq.~(\ref{eq:itmd}. We have obtained the first 
numerical implementation of this formalism, and the first predictions for forward di-jets at the LHC, a process
which is particularly interesting from small-$x$ point of view. Our results for the nuclear modification factors
in p+Pb vs p+p collisions confirm the conclusions obtained in \cite{vanHameren:2014lna} in the HEF framework,
that for nearly back-to-back jets, non negligible effects of gluon saturation are to be expected as
one goes from p+p to p+Pb collisions.

\section*{Acknowledgements}
K.K acknowledges support by Narodowe Centrum Nauki with Sonata Bis grant DEC-2013/10/E/ST2/00656. The article is based on paper written with P. Kotko, C. Marquet, E. Petreska, S. Sapeta, A. van Hameren


\begin{thebibliography}{99}


\bibitem{Deak:2009xt}
  M.~Deak, F.~Hautmann, H.~Jung and K.~Kutak,
  JHEP {\bf 0909} (2009) 121
  doi:10.1088/1126-6708/2009/09/121
  [arXiv:0908.0538 [hep-ph]].



\bibitem{Deak:2009ae}
  M.~Deak, F.~Hautmann, H.~Jung and K.~Kutak,
  arXiv:0908.1870 [hep-ph].




\bibitem{Deak:2010gk}
  M.~Deak, F.~Hautmann, H.~Jung and K.~Kutak,
  arXiv:1012.6037 [hep-ph].

\bibitem{Deak:2011ga}
  M.~Deak, F.~Hautmann, H.~Jung and K.~Kutak,
  Eur.\ Phys.\ J.\ C {\bf 72} (2012) 1982
  doi:10.1140/epjc/s10052-012-1982-5
  [arXiv:1112.6354 [hep-ph]].


\bibitem{Deak:2011gj}
  M.~Deak, F.~Hautmann, H.~Jung and K.~Kutak,
  arXiv:1112.6386 [hep-ph].


\bibitem{Kutak:2012rf}
  K.~Kutak and S.~Sapeta,
  Phys.\ Rev.\ D {\bf 86} (2012) 094043
  doi:10.1103/PhysRevD.86.094043
  [arXiv:1205.5035 [hep-ph]].



\bibitem{Marquet:2007vb}
  C.~Marquet,
  Nucl.\ Phys.\ A {\bf 796} (2007) 41.


\bibitem{Gribov:1984tu}
  L.~V.~Gribov, E.~M.~Levin and M.~G.~Ryskin,
  Phys.\ Rept.\  {\bf 100} (1983) 1.
  
  
\bibitem{Albacete:2010pg}
  J.~L.~Albacete and C.~Marquet,
  Phys.\ Rev.\ Lett.\  {\bf 105} (2010) 162301.
  
  \bibitem{Stasto:2011ru}
  A.~Stasto, B.~-W.~Xiao and F.~Yuan,
  Phys.\ Lett.\ B {\bf 716} (2012) 430.
  
\bibitem{Lappi:2012nh}
  T.~Lappi and H.~Mantysaari,
  Nucl.\ Phys.\ A {\bf 908} (2013) 51.
  
\bibitem{Gelis:2010nm}
  F.~Gelis, E.~Iancu, J.~Jalilian-Marian and R.~Venugopalan,
  Ann.\ Rev.\ Nucl.\ Part.\ Sci.\  {\bf 60} (2010) 463.

\bibitem{Albacete:2014fwa} 
  J.~L.~Albacete and C.~Marquet,
  Prog.\ Part.\ Nucl.\ Phys.\  {\bf 76} (2014) 1.

\bibitem{Adare:2011sc}
  A.~Adare {\it et al.}  [PHENIX Collaboration],
  Phys.\ Rev.\ Lett.\  {\bf 107} (2011) 172301.

\bibitem{Braidot:2010zh}
  E.~Braidot [STAR Collaboration],
  arXiv:1005.2378 [hep-ph].
  
  
  \bibitem{Kotko:2015ura}
  P.~Kotko, K.~Kutak, C.~Marquet, E.~Petreska, S.~Sapeta and A.~van Hameren,
  JHEP {\bf 1509} (2015) 106 
  
  
\bibitem{Mueller:2012uf}
  A.~H.~Mueller, B.~-W.~Xiao and F.~Yuan,
  Phys.\ Rev.\ Lett.\  {\bf 110} (2013) 082301.
  
\bibitem{Mueller:2013wwa}
  A.~H.~Mueller, B.~-W.~Xiao and F.~Yuan,
  Phys.\ Rev.\ D {\bf 88} (2013) 114010.
  
\bibitem{vanHameren:2014ala}
  A.~van Hameren, P.~Kotko, K.~Kutak and S.~Sapeta,
  Phys.\ Lett.\ B {\bf 737} (2014) 335.

\bibitem{Kutak:2014wga}
  K.~Kutak,
  Phys.\ Rev.\ D {\bf 91} (2015) no.3,  034021.

\bibitem{Ciafaloni:1987ur}
  M.~Ciafaloni,
  Nucl.\ Phys.\ B {\bf 296} (1988) 49.
  
\bibitem{Catani:1989sg}
  S.~Catani, F.~Fiorani and G.~Marchesini,
  Nucl.\ Phys.\ B {\bf 336} (1990) 18.
  
 \bibitem{Catani:1989yc}
  S.~Catani, F.~Fiorani and G.~Marchesini,
  Phys.\ Lett.\ B {\bf 234} (1990) 339.

\bibitem{vanHameren:2016ftb}
  A.~van Hameren, P.~Kotko, K.~Kutak, C.~Marquet, E.~Petreska and S.~Sapeta,
  JHEP {\bf 1612} (2016) 034

\bibitem{Catani:1990eg}
  S.~Catani, M.~Ciafaloni and F.~Hautmann,
  Nucl.\ Phys.\ B {\bf 366} (1991) 135.

\bibitem{Angeles-Martinez:2015sea}
  R.~Angeles-Martinez {\it et al.},
  Acta Phys.\ Polon.\ B {\bf 46} (2015) no.12,  2501
  doi:10.5506/APhysPolB.46.2501
  [arXiv:1507.05267 [hep-ph]].



\bibitem{Dominguez:2011wm}
  F.~Dominguez, C.~Marquet, B.~W.~Xiao and F.~Yuan,
  Phys.\ Rev.\ D {\bf 83} (2011) 105005
  doi:10.1103/PhysRevD.83.105005
  [arXiv:1101.0715 [hep-ph]].

\bibitem{vanHameren:2016kkz}
  A.~van Hameren,
  arXiv:1611.00680 [hep-ph].
\bibitem{KotkoLxJet}
P. Kotko. LxJet, the code is available at LxJet.html. http://annapurna.ifj.edu.pl/~pkotko/

\bibitem{vanHameren:2014lna}
  A.~van Hameren, P.~Kotko, K.~Kutak, C.~Marquet and S.~Sapeta,
  Phys.\ Rev.\ D {\bf 89} (2014) no.9,  094014
  doi:10.1103/PhysRevD.89.094014
  [arXiv:1402.5065 [hep-ph]].

\end{thebibliography}
\end{document}